# Visualizing the evolution from the Mott insulator to a charge ordered insulator in lightly doped cuprates

Peng Cai,[1] Wei Ruan,[1] Yingying Peng,[2] Cun Ye,[1] Xintong Li,[1] Zhenqi Hao,[1] Xingjiang Zhou,[2,5] Dung-Hai Lee,[3,4] Yayu Wang[1,5†]

[1]*State Key Laboratory of Low Dimensional Quantum Physics, Department of Physics, Tsinghua University, Beijing 100084, P.R. China*

[2]*Beijing National Laboratory for Condensed Matter Physics, Institute of Physics, Chinese Academy of Sciences, Beijing 100190, P. R. China*

[3]*Department of Physics, University of California at Berkeley, Berkeley, CA 94720.*

[4]*Materials Sciences Division, Lawrence Berkeley National Laboratory, Berkeley, CA 94720.*

[5]*Innovation Center of Quantum Matter, Beijing 100084, P.R. China*

[†] Email: yayuwang@tsinghua.edu.cn

A central question in the high temperature cuprate superconductors is the fate of the parent Mott insulator upon charge doping. Here we use scanning tunneling microscopy to investigate the local electronic structure of lightly doped cuprate in the antiferromagnetic insulating regime. We show that the doped charge induces a spectral weight transfer from the high energy Hubbard bands to the low energy in-gap states. With increasing doping, a V-shaped density of state suppression occurs at the Fermi level, which is accompanied by the emergence of checkerboard charge order. The new STM perspective revealed here is the cuprates first become a charge ordered insulator upon doping. Subsequently, with further doping, Fermi surface and high temperature superconductivity grow out of it.

High temperature superconductivity in the cuprates is widely believed to originate from adding charge carriers into an antiferromagnetic (AF) Mott insulator (*1*). Elucidating the properties of the doped Mott insulator is among the most crucial issues concerning the mechanism of superconductivity. From the electronic structure point of view, the key question is how the large Mott-Hubbard gap, or more precisely the charge transfer gap, evolves into the *d*-wave superconducting (SC) gap upon charge doping. This has turned out to be a formidable challenge, both theoretically and experimentally, due to the presence of strong AF fluctuations and electron correlations. A major obstacle lying between the parent Mott insulator and the optimally doped cuprate is the pseudogap phase, which exhibits a normal state gap as revealed by early spectroscopic studies (*2-4*). More recently, imaging and diffraction techniques show that electrons in the pseudogap phase have strong propensity towards charge (*5-16*) or spin order (*5, 17-19*). Currently neither the gap-like density of state (DOS) suppression nor the charge/spin density wave is well understood (*4*).

Most previous experiments on the pseudogap phase focused on the underdoped regime with finite transition temperature ($T_c$), and aimed to address its relationship with the SC phase by probing the strength of the two orders across $T_c$ (*4*). Recent scanning tunneling microscopy (STM) and x-ray spectroscopy experiments provide increasing evidence that the charge order associated with the pseudogap compete with superconductivity because its feature gets suppressed as the sample enters the SC phase below $T_c$ (*4, 11, 16*). However, due to the lack of spectroscopic data spanning the energy range of both the pseudogap and charge transfer gap, less is known about lightly doped, non-SC regime standing next to the parent Mott insulator. Despite recent progresses in STM experiments on severely underdoped cuprates (*20*), questions such as how the pseudogap evolves within the large charge transfer gap as a function of doping, what is the relation (if any) between the heavily discussed charge order and pseudogap, and how superconductivity emerges out of the pseudogap are still open.

To address these questions, here we carry out STM investigations on lightly doped $Bi_2Sr_{2-x}La_xCuO_{6+\delta}$ (La-Bi2201) in the AF insulating regime. La-Bi2201 is an ideal cuprate system which not only has a well-cleaved surface, but also can be doped towards the Mott insulating limit by varying the La and O contents (*21-23*). Figure 1A displays the schematic phase diagram of La-Bi2201, in which long-range AF order extends to nominal hole density $p \sim 0.10$ and superconductivity emerges with further doping. The two samples studied in this work have hole density $p = 0.03$ and 0.07 (the method to estimate the hole density is described in ref. *21-23* and the reference therein), which are in the AF-ordered, strongly insulating regime. Although it is technically challenging to perform STM experiments on insulators, we have developed reliable methodology for high resolution topography and large bias spectroscopy on insulating curpates, as demonstrated by previous STM results on pristine $Ca_2CuO_2Cl_2$ (CCOC) Mott insulator (*24*). More experimental details are described in the Methods session.

Figure 1B shows the STM topography of a cleaved $p = 0.03$ sample acquired at $T = 77$ K, and the setup parameters are bias voltage $V = -1.2$ V and tunneling current $I = 2$ pA. The structural supermodulation and the Bi atoms on the top surface can be clearly resolved. Fig. 1C displays spatially resolved tunneling spectra $dI/dV(r, V)$, which is roughly proportional to the local electron DOS, collected along the red line in the topography. The $dI/dV$ curves have pronounced spatial variations due to the electronic inhomogeneity in underdoped cuprates, but overall they exhibit highly systematic spectral evolutions. In Fig. 1D we display three representative $dI/dV$ curves taken in this area, and below we will describe their features one by one.

In the black $dI/dV$ curve in Fig. 1D, there is no DOS at all for the energy range from -0.2 eV to +1.5 eV. Such a lineshape indicates the existence of a large energy gap with a size of ~ 1.7 eV, which is close to the expected charge transfer gap value for the parent cuprate (*1*). To get more insight into this energy gap, in Fig. 1E we compare this spectrum with that of pristine CCOC (*24*), a well-defined parent Mott insulator. The two curves show very similar behavior, except that the charge transfer gap size in CCOC is $\Delta_{CTG} \sim 2.2$ eV, about 0.5 eV larger than that in La-Bi2201. By rescaling the energies with their respective $\Delta_{CTG}$, we find that the two spectra collapse to an identical one (Fig. 1E inset), confirming that the spectrum of La-Bi2201 is indeed taken on a parent phase with negligible doping. The blue curve in Fig. 1D represents the most common type of spectrum in this sample, which shows a broad in-gap state emerging within the charge transfer gap. This in-gap state extends all the way from the lower Hubbard band (LHB) to the upper Hubbard band (UHB), but at $E_F$ the DOS is still zero. The overall lineshape bears strong resemblance to that taken on individual hole-type dopant at the Ca cite of CCOC (Fig. S1), suggesting that this spectrum is taken in an area with very dilute hole concentration. The red curve in Fig. 1D occurs in a small portion of sample, in which the in-gap state becomes more pronounced. Due to the rapid growth of low energy DOS at both sides of $E_F$, the spectrum has a gap-like lineshape centered around $E_F$. An important trend revealed by the comparison of the three curves is that with the increase of low energy spectral weight within the charge transfer gap, the spectral weight at the UHB decreases. The low energy in-gap state is thus induced by spectral weight transfer from the high energy Hubbard bands.

To visualize the spatial distribution of the DOS, we take conductance maps $dI/dV(r, V)$ at different biases on the area marked by the yellow dashed square in Fig. 1B. At zero bias, which corresponds to the Fermi level, the conductance map (Fig. 2B) shows zero DOS everywhere. This indicates the absence of low energy gapless single electron excitations in this sample that is consistent with the strongly insulating transport behavior (*21, 23*). At small biases (Fig. 2A and 2C), the DOS is zero in the nanometer-sized dark areas corresponding to the undoped Mott insulator phase, but becomes finite in the bright areas due to the emergence of low energy in-gap states at slightly higher doping. At high biases associated with the UHB (Fig. 2D), the spectral weight is higher in the parent Mott phase but is lower for the areas with the in-gap state, showing an anti-correlation to the low bias ones. This is a direct illustration of the spectral weight transfer process. In contrast, the low bias maps are positively correlated, indicating that the low energy states on both sides of $E_F$ emerge simultaneously (see Fig. S2 for more DOS maps in this sample).

To understand how the electronic structure evolves with further hole doping, we next move to the $p = 0.07$ sample. Fig. 3A shows a series of representative spectra taken at $T = 5$ K on the locations indicated by the topography in the inset. These spectra still show smooth evolution with continuous spectral weight transfer from high energy to low energy. Fig. 3B displays the spectra over a smaller energy window, which illustrates the narrowing of the energy gap near $E_F$. Due to the extreme particle-hole asymmetry near $E_F$, the gap edge is more pronounced at the positive bias. Therefore here we use the energy scale showing the maximum within the CTG to roughly characterize the gap size. The gap size map in Fig. 3B inset shows that the majority of the sample has a gap amplitude in the range of 80 to 200 meV (*20, 25*). The DOS at $E_F$ is still zero everywhere (Fig. 3D), indicating the absence of gapless excitations. In addition, low bias conductance maps (Fig. 3C, 3E and 3F) show that the region with pronounced in-gap states dominate, as can be seen from the patches of bright areas that occupy the majority of the sample. Similar to the *p* = 0.03 sample, the spatial distribution of the spectral weight is positively correlated between the positive and negative low biases, and anti-correlated between low and high biases. Moreover, the gap feature near $E_F$ becomes narrower with increasing in-gap spectral weight, hence the local hole doping. Meanwhile, the DOS becomes more particle-hole symmetric around $E_F$, most likely due to the weakening of electron correlation effect as the sample moves away from the Mott insulator limit (*26, 27*).

A closer examination of the *dI/dV* maps reveals another highly intriguing phenomenon. On the bright areas of the low bias conductance maps (Fig. 3C, E and F), a checkerboard-like pattern emerges. The periodicity of the checkerboard is bias-independent from -25 mV to +100 mV (see more DOS maps in Fig. S3), indicating a static charge order. The charge order pattern can be better observed in the current map *I* (r, V) shown in Fig. 4A taken at *V* = +100 mV, which by definition is the integral of the *dI/dV* maps from 0 to + 100 mV. Based on the Fourier transform of the current map (Fig. 4B), the orientation of the checkerboard is shown to be along the Cu-O bond direction. A line cut along this direction shows that the checkerboard wavevector is around 1/4 of the Cu-Cu lattice wavevector $2\pi/a_0$ (Fig. 4C), corresponding to a real space wavelength of $\sim 4a_0$ ($a_0$ is the distance between neighboring Cu atoms). The full width at half maximum of the charge order peak is $\sim 0.08/a_0$, equivalent to a real-space correlation length $\sim 12\ a_0$. The charge order is thus rather short-ranged, which can also be estimated directly from the size of the patches showing the checkerboard (see Fig. S4 for the same charge order pattern observed on another *p* = 0.07 sample).

More interestingly, there is a close correlation between the real space charge order and the local electronic structure. As shown in Fig. 3E, the *dI/dV* spectrum taken in the dark area without checkerboard order is close to that of the parent Mott insulator (black curve in Fig. 3A), showing a large charge transfer gap with a very weak and broad in-gap state. In contrast, the bright area with well-defined checkerboard order shows the sharp V-shaped gap near $E_F$ (red curve in Fig. 3A), and the high energy spectral weight is so much reduced that the UHB cannot be clearly resolved up to 2 V. In the crossover area lying between these two limits where the charge puddles start to nucleate, the spectrum shows an intermediate behavior with broad but pronounced in-gap state. The checkerboard order thus emerges gradually from the Mott insulator limit and becomes more pronounced with increasing hole doping.

The STM results reported here touch upon several important issues regarding the electronic structure of doped Mott insulators in cuprates. Our large bias $dI/dV$ spectroscopy reveals directly that the main effect of hole doping is to cause a spectral weight transfer from the high energy Hubbard bands to the low energy state near $E_F$. This can be explained by the fact that doped holes create electron vacancies so that electron injection no longer needs to overcome the onsite Coulomb repulsion energy, which effectively transfers the spectral weight from the "UHB". Moreover, as a function of doping a continuous family of spectral shapes with zero DOS at $E_F$ and growing spectral weight within the charge transfer gap are observed. Because the evolution is continuous we should regard all of them as exhibiting the "pseudogap". However, among this family of spectral shapes the DOS suppression around $E_F$ is characterized by a range of energy scales, starting from around 500 meV for $p = 0.03$ to as small as 80 meV for $p = 0.07$. Interestingly, at the high doping end a relatively sharp V-shaped DOS suppression (red curve in Fig. 3B) reminiscent of the commonly observed pseudogap emerges.

Next we will address the origin of the checkerboard charge order. Previous proposals for the checkerboard order can be classified into two groups, the momentum-space picture and the real-space picture. The former one attributes the charge density modulation to FS nesting along the antinodal region of the Brillouin zone, such as charge density wave (*28-30*), SC order with a finite Cooper pair momentum (*31, 32*), and impurity potential scattering (*33*). The $p = 0.07$ sample studied here is still in the highly insulating regime, and spectroscopically there is no gapless excitations. Since the $dI/dV$ curve for this doping continuously evolves from the parent Mott insulator upon spectral weight transfer from the Hubbard bands, we conclude the charge ordered insulator state grows out of the parent Mott insulator rather than a metallic state. Therefore, we believe the real-space pictures such as a charge order accompanying the spin density wave (*34*) or stripe order (*35*) are more relevant. The observation that checkerboard emerges gradually with increasing spectral weight transfer from the UHB suggests that it maybe already hidden in the Mott insulating state. As a sufficient amount of holes are doped, the appearance of charge order with $4a_0$ periodicity is due to both the underlying spin configuration (*18, 34*) and the lattice pinning effect (*36, 37*). Particularly, it was proposed that it is energetically favorable for the doped holes to bind locally in case of Cu-O bond contraction, hence forming short-ranged $4a_0$ checkerboard pattern (*38*).

Another important issue is the relationship between the charge order and superconductivity. Recent STM experiments on cuprates with coexisting checkerboard and SC phases suggest that they compete with each other and superconductivity wins over the charge order at temperatures below $T_c$ (*16*). Our experiment addresses this issue through a complementary approach, in which we vary the hole concentration towards the SC phase boundary at the ground state from parent Mott insulator. We show that when superconductivity is completely suppressed by reduced doping the resulting state is checkerboard charge ordered. This raises the possibility that the charge order could actually be a periodic pattern of localized Cooper pairs, or crystal of Cooper pairs (*39*). Upon doping superconductivity emerges when the localized Cooper pairs become itinerant. The FS pocket observed by quantum oscillation (*40*) when superconductivity is suppressed by strong magnetic field then manifests the metallic state grows out of the charge ordered state.

Our work resolves another key controversy regarding the doped Mott insulators, i.e., which phase appears first upon hole doping. Here we demonstrate that from the STM perspective the charge order, rather than superconductivity, is the first electronic ordered state that emerges by doping the parent compound. This is consistent with the density matrix renormalization group calculations performed on *t-J* or Hubbard models which showed that doped Mott insulator is always charge ordered (*41, 42*). Thus the metallic, hence the SC, state of the cuprates actually grows out of a charge ordered insulating state. A critical issue that deserves further investigation is whether the Cooper pairs form on the FS of the metallic state that grows out of the charge ordered state, or they already form but just being localized in the charge ordered state.

**Acknowledgments:** We thank T.K. Lee, N. Trivedi, F. Wang, Z.Y. Weng, T. Xiang, and G.M. Zhang for helpful discussions. This work is supported by the NSFC (11190022, 11334010 and 11374335) and MOST of China (2011CB921703, 2011CBA00110, 2015CB921000), and the Chinese Academy of Sciences (XDB07020300). DHL was supported by the U.S. Department of Energy, Office of Science, Basic Energy Sciences, Materials Sciences and Engineering Division, grant DE-AC02-05CH11231.

Figure Captions

**Fig. 1** Electronic structure of lightly hole doped La-Bi2201 near the Mott insulator limit. (**A**) Schematic electronic phase diagram of La-Bi2201 showing the antiferromagnetic insulator, superconductor and pseudogap phases. The black arrows represent the two samples studied here with hole density $p$ = 0.03 and 0.07, respectively. (**B**) Atomically resolved topography of the $p$ = 0.03 sample acquired at $T$ = 77 K with bias voltage $V$ = -1.2 V and tunneling current $I$ = 2 pA over an area of 250×250 Å$^2$. (**C**) Spatially resolved large bias range $dI/dV$ spectra taken along the red line in (B). Each spectrum is shifted vertically for clarity. (**D**) Three representative spectra taken on the $p$ = 0.03 sample. The black curve shows the charge transfer gap of the parent state, the blue and red curves exhibit broad in-gap states emerging within the charge transfer gap. (**E**) Comparison of the charge transfer gap in La-Bi2201 (blue) and pristine CCOC (black). The inset shows the identical lineshape of the two spectra when they are normalized.

**Fig. 2.** DOS map of the $p$ = 0.03 sample measured at varied biases. (**A-D**) $dI/dV$ maps taken on the field of view marked by the yellow dashed square in Fig. 1(B) at $V$ = -0.1, 0, +0.2 and +1.6 V respectively. The bright areas in A and C indicate the existence of low energy in-gap states, and the dark areas are the undoped Mott

insulator phase. The anti-correlation between the low bias (A, C) and large bias (D) maps demonstrates the spectral weight transfer from the UHB to low energy excitations in doped Mott insulator. The complete darkness in (B) reveals the absence of DOS at $E_F$ for the whole sample.

**Fig. 3.** The electronic structure evolution and checkerboard charge order in the $p$ = 0.07 sample. (**A**) Five representative $dI/dV$ spectra acquired at $T$ = 5 K on different locations of the $p$ = 0.07 sample with setup parameters $V$ = -1.2 V and $I$ = 4 pA. The spectrum evolves systematically from close to the Mott insulator state to broad in-gap state and then to the V-shaped DOS suppression. (Inset) Topography measured at $V$ = -0.5 V and $I$ = 2 pA over an area of 330×330 Å$^2$. The cross and numbers denote the locations where the spectra are taken. (**B**) Low bias $dI/dV$ spectra measured on the same locations, which reveal the evolution of the low energy gap features. (Inset) Gap map measured on the same area as the inset of (A). (**C-G**) $dI/dV$ maps measured at varied biases. In addition to the correlation relation between spectral weight at different biases, a checkerboard charge order can be clearly observed in the low bias maps (C, E, F). The DOS at $E_F$ is still zero everywhere (D), consistent with the insulating behavior.

**Fig. 4.** The wavevector of the checkerboard charge order. (**A**) The tunneling current map $I(r, V)$ measured at $V$ = +100 mV, which is equivalent to the integrated $dI/dV$ maps from 0 to +100 mV, on the same area as the inset of Fig. 3A. (**B**) Fourier transform of the current map in (A), revealing the wavevector of the lattice and charge order respectively. (**C**) The line-cut along the Cu-O bond direction in (B) shows a clear charge order peak with wavevector around 1/4 of the Cu-Cu lattice wavevector $2\pi/a_0$, corresponding to a $4a_0$ real-space periodicity for the checkerboard. The red line is a Gaussian fit for the charge order peak.

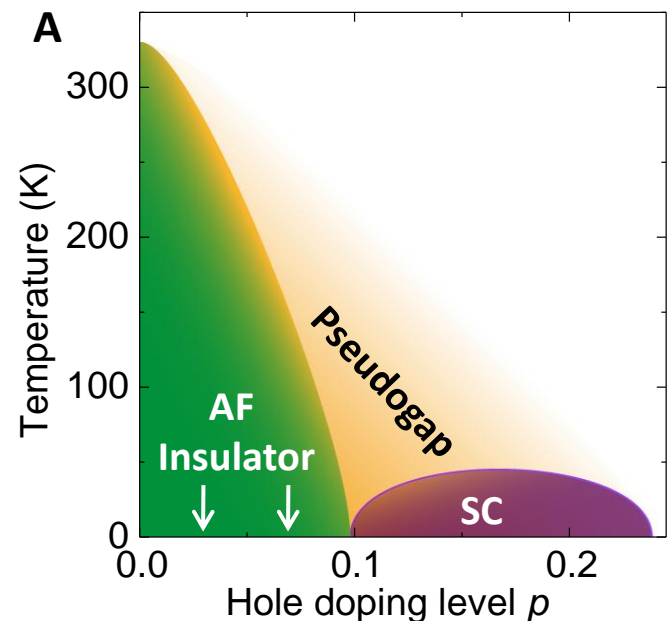
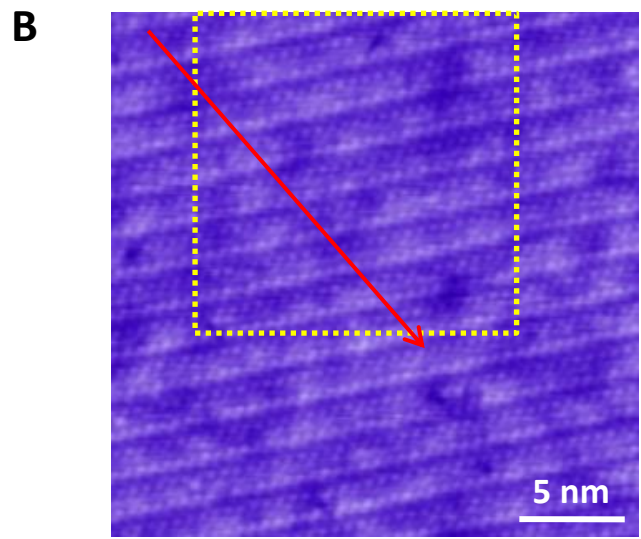
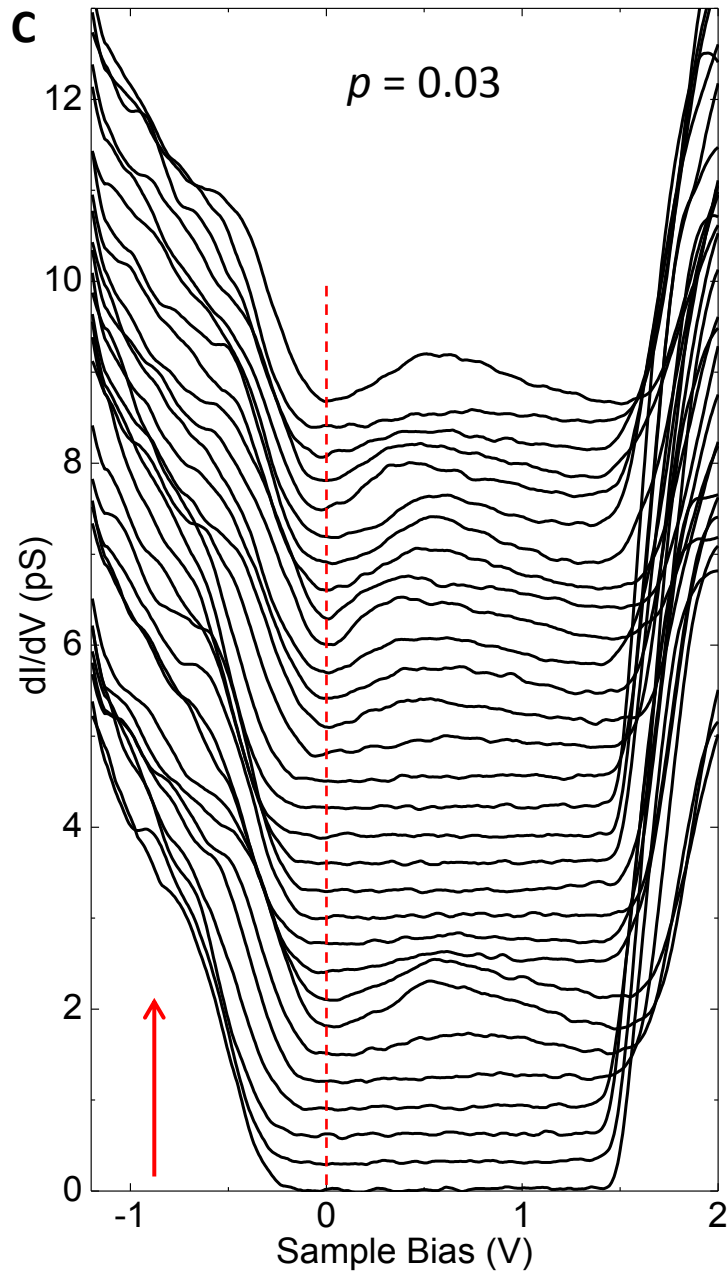
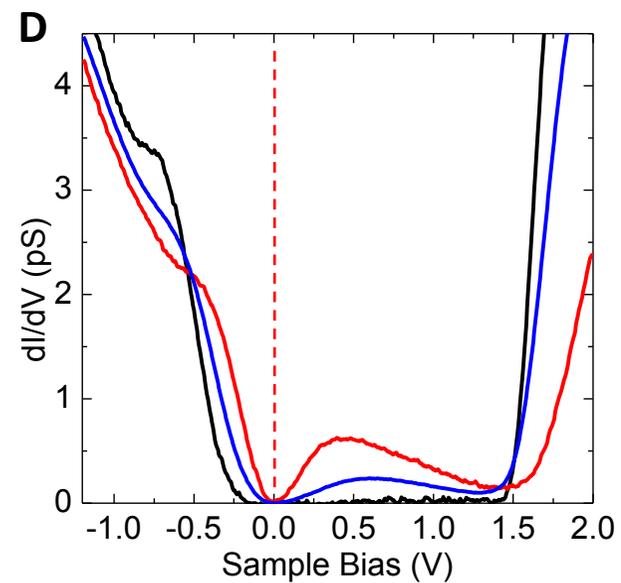
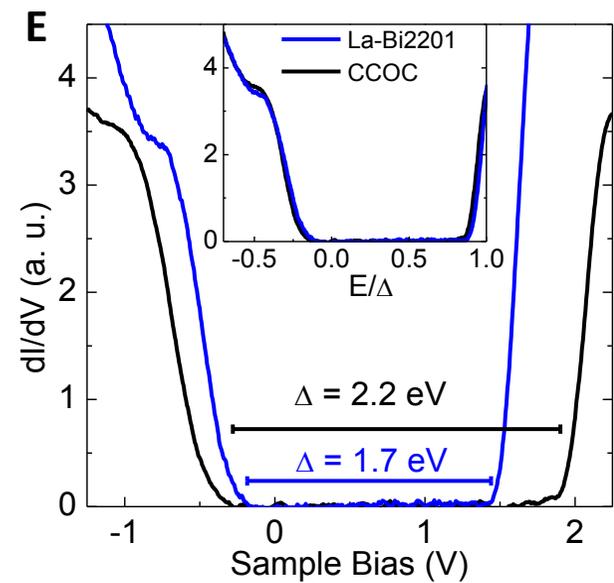

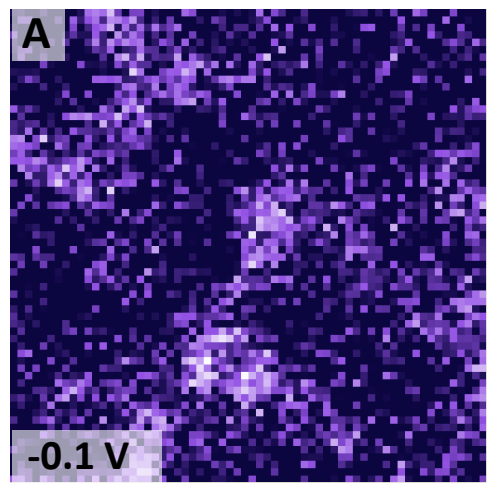 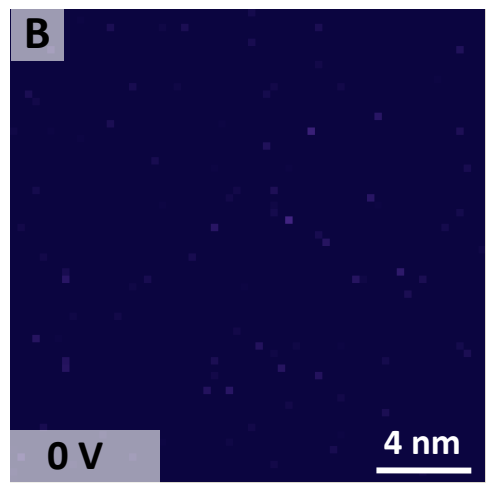 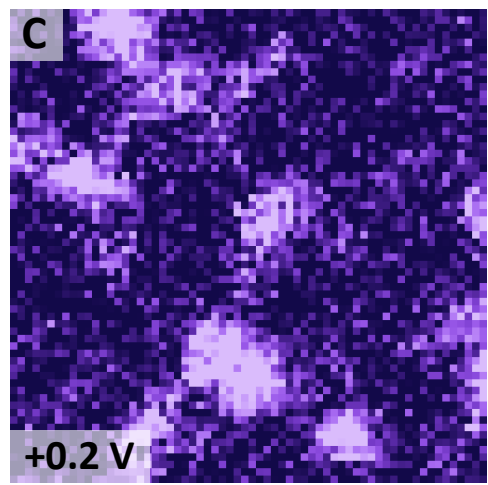 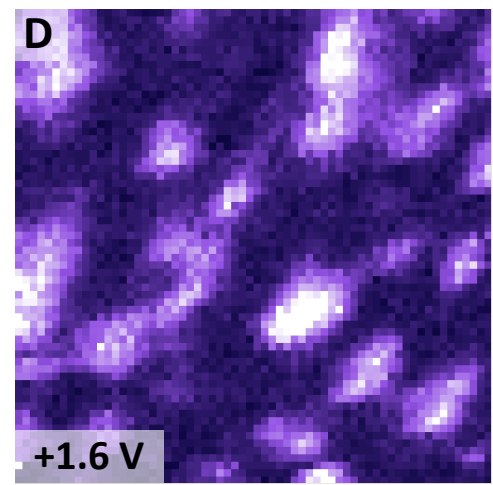

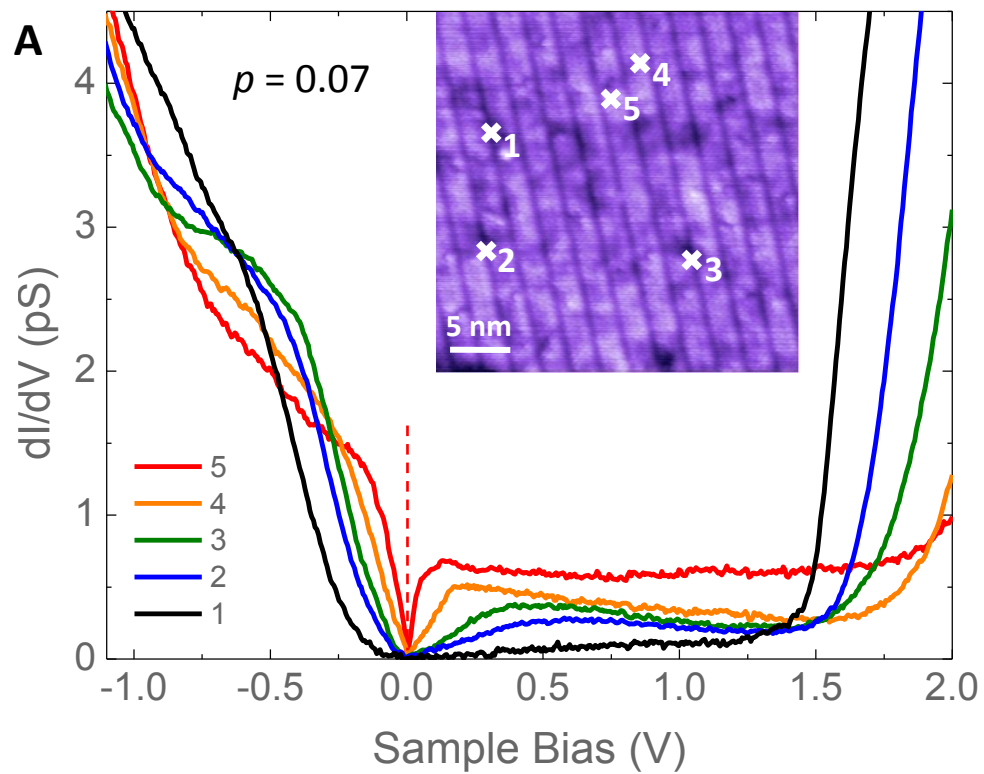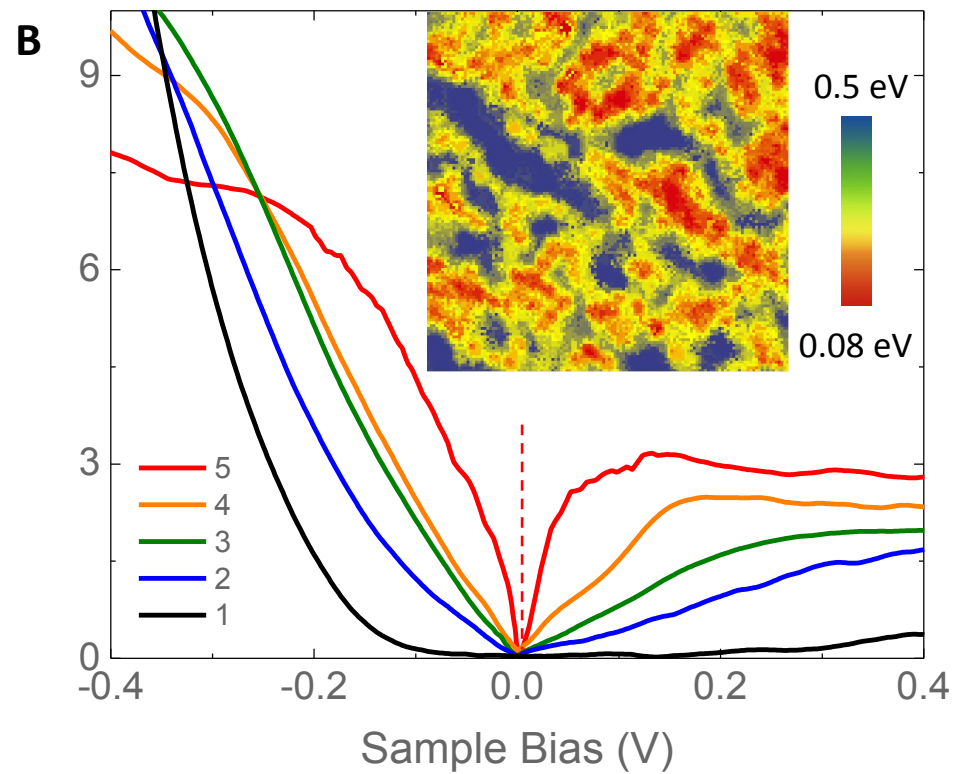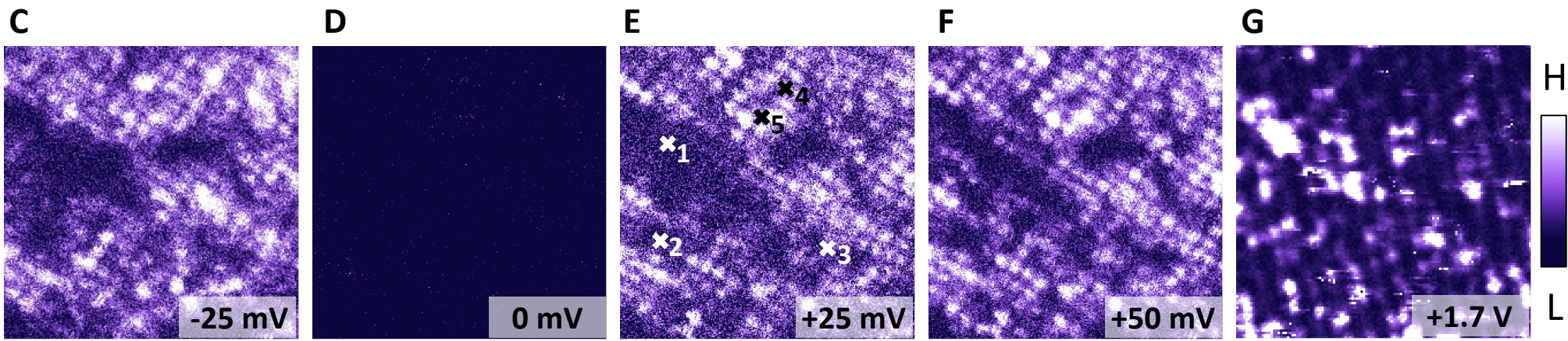

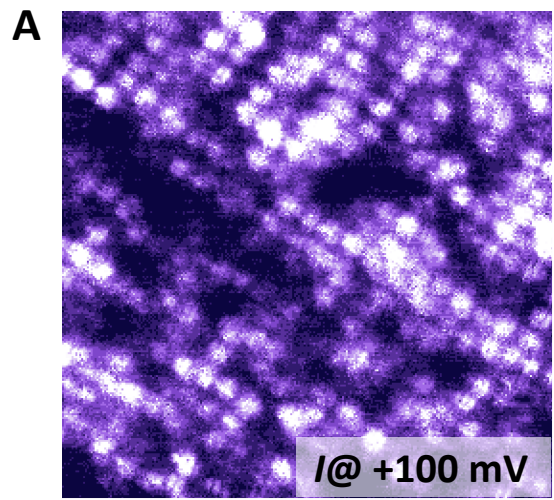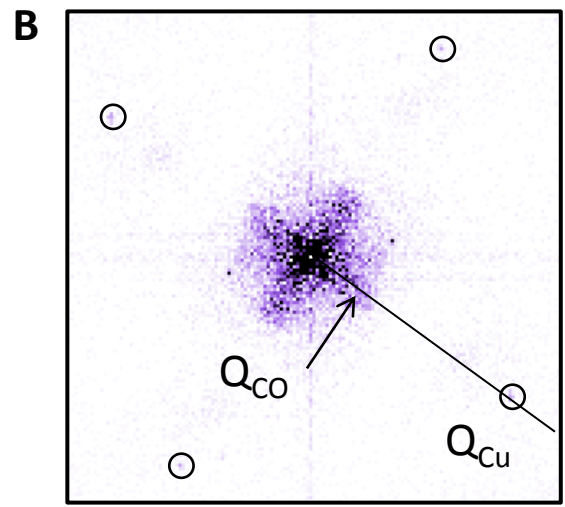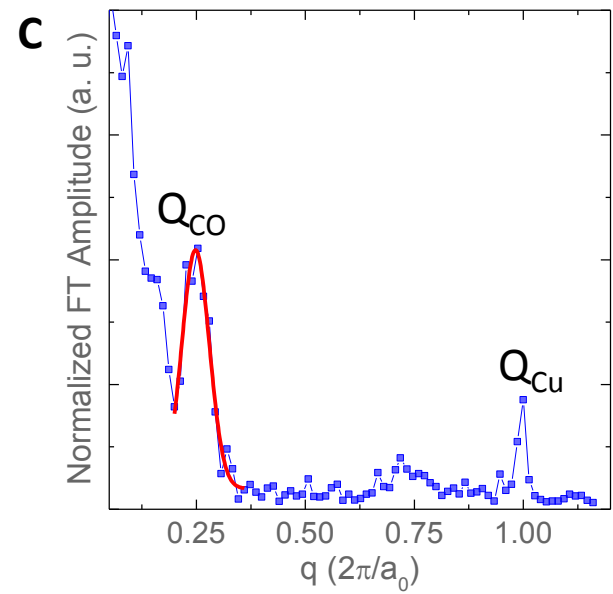

Supporting Online Materials for

# Visualizing the evolution from the Mott insulator to a charge ordered insulator in lightly doped cuprates

Peng Cai,[1] Wei Ruan,[1] Yingying Peng,[2] Cun Ye,[1] Xintong Li,[1] Zhenqi Hao,[1] Xingjiang Zhou,[2,5] Dung-Hai Lee,[3,4] Yayu Wang[1,5†]

[1]*State Key Laboratory of Low Dimensional Quantum Physics, Department of Physics, Tsinghua University, Beijing 100084, P.R. China*

[2] *Beijing National Laboratory for Condensed Matter Physics, Institute of Physics, Chinese Academy of Sciences, Beijing 100190, P. R. China*

[3]*Department of Physics, University of California at Berkeley, Berkeley, CA 94720.*

[4]*Materials Sciences Division, Lawrence Berkeley National Laboratory, Berkeley, CA 94720.*

[5]*Innovation Center of Quantum Matter, Beijing 100084, P.R. China*

[†] Email: yayuwang@tsinghua.edu.cn

## Contents:

**Experimental methods**

**Supporting materials text**

**Figure S1 to S4**

**References**

## I. Experimental methods

Single crystals of $Bi_2Sr_{2-x}La_xCuO_{6+\delta}$ (La-Bi2201) are grown by the travelling solvent floating zone method. The sample growth and characterization methods have been reported previously (*1*). The STM experiments are performed using a low temperature ultrahigh vacuum STM system manufactured by CreaTec. The La-Bi2201 single crystal is cleaved *in situ* at $T = 77$ K and then transferred immediately into the STM sample stage. For the strongly insulating $p = 0.03$ sample, no tunneling current can be detected at $T = 5$ K and the STM tip will crash into the sample at such low temperature. Therefore, all the STM data on this sample are acquired at $T = 77$ K, when the thermally activated charge carriers allow a small tunneling current to flow through the sample. For the less insulating $p = 0.07$ sample, it is possible to perform STM experiments at lower temperature so the data reported here are acquired at $T = 5$ K. The STM topography is taken in the constant current mode, and the *dI/dV* spectra are collected using a standard lock-in technique with modulation frequency $f = 523$ Hz. To obtain high quality STM data over a large bias range at high temperatures, it is crucial to have a clean and stable tip. We use an electrochemically etched tungsten tip, which is treated and calibrated carefully *in situ*, as described in details in our previous paper (*2*).

## II. Similar *dI/dV* spectra in lightly doped La-Bi2201 and CCOC

The *dI/dV* spectra in the very lightly doped La-Bi2201 with $p = 0.03$ are highly analogous to that observed in pristine CCOC with sparse hole-type defects. As shown in Fig. S1A here, the red curve shows a broad in-gap state that extends all the way from the lower Hubbard band to the upper Hubbard band. The DOS at $E_F$ remains zero, and the spectral weight of the UHB is strongly suppressed compared to the undoped Mott insulator phase (the black curve). The red curve in Fig. S1B is taken on a Ca-site defect in pristine CCOC Mott insulator. Although the exact nature of the Ca-site defect is unknown, it is most likely due to Ca vacancy or Na substitution. In either case the defect will introduce hole-type carrier into the Mott insulator. Its overall spectral lineshape show remarkable resemblance to the red curve in Fig. S1A. The strong similarity between the spectra in two totally different cuprate systems indicate that the broad in-gap state, the suppression of the UHB spectral weight, and the zero DOS at $E_F$ are universal features of the cuprate parent Mott insulator doped with very dilute holes. Another similarity between the two systems is the strong spatial localization of the in-gap state. In CCOC, the broad in-gap state dies out at 4-5 lattice distances, or around 2 nm, from the defect center. For the La-Bi2201, the DOS map shown in Fig. 2C of the main text reveals that the puddles with broad in-gap state have a typical size of 2-4 nm. The spatial localization is another character of the electronic states induced by hole-type carriers in underdoped cuprate close to the Mott insulator limit.

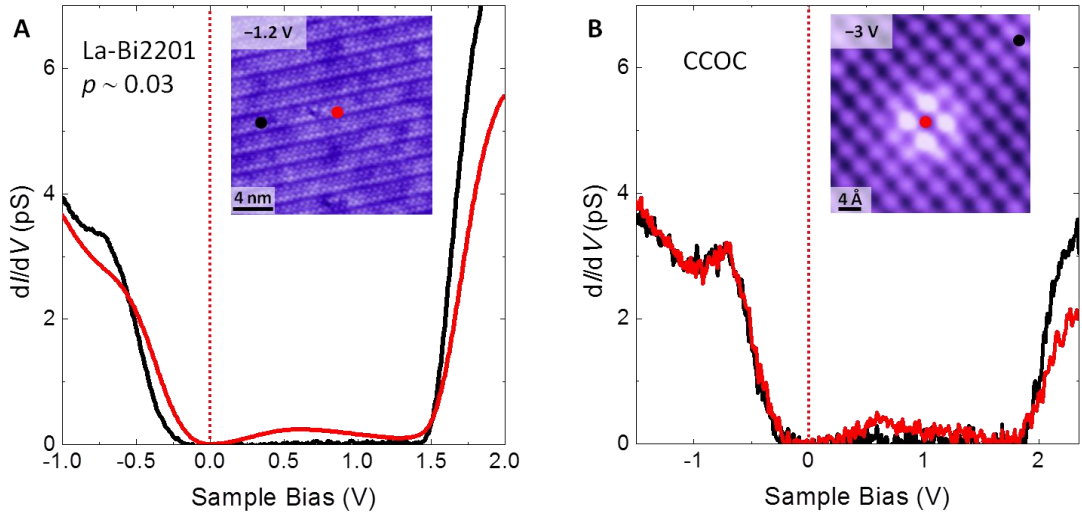

Fig. S1: (A) Representative $dI/dV$ spectra in the very lightly doped La-Bi2201 with $p = 0.03$. The setup parameters are $V$ = -1.2V and $I$ = 4pA. (B) $dI/dV$ spectra measured on Ca-cite defect (red) and defect-free cite (black) of pristine CCOC. The color spots in the inset topographic images indicate the position where the spectra are acquired.

### III. Detailed DOS maps of the $p = 0.03$ sample

In Fig. 2 of the main text we display four representative DOS maps of the $p = 0.03$ samples with sample bias $V$ = -0.1, 0, 0.2, and 1.6 V respectively. In this session we show detailed raw data (in gray scale) of the DOS maps with more bias voltages, in which the evolution and correlation of the electronic states can be thoroughly visualized. The $dI/dV$ maps in Fig. S2 are taken on the same area of the $p = 0.03$ sample as that shown in Fig. 2 of the main text. The maps from -0.4 V to +1.0 V clearly show the absence of gapless excitations at $E_F$ and the emergence of the low energy in-gap state at finite biases. All these maps show similar spatial patterns indicative of positive correlation of the low energy states. The brighter features at negative bias reflect the particle-hole asymmetry in the lightly doped cuprate, in which the occupied states have larger DOS than the empty states (*3*). The contrast reverses above +1.4 V, indicating the spectral weight transfer from the high energy UHB to low energy in-gap states. Above +1.8 V, the structural supermodulation starts to appear in the DOS maps.

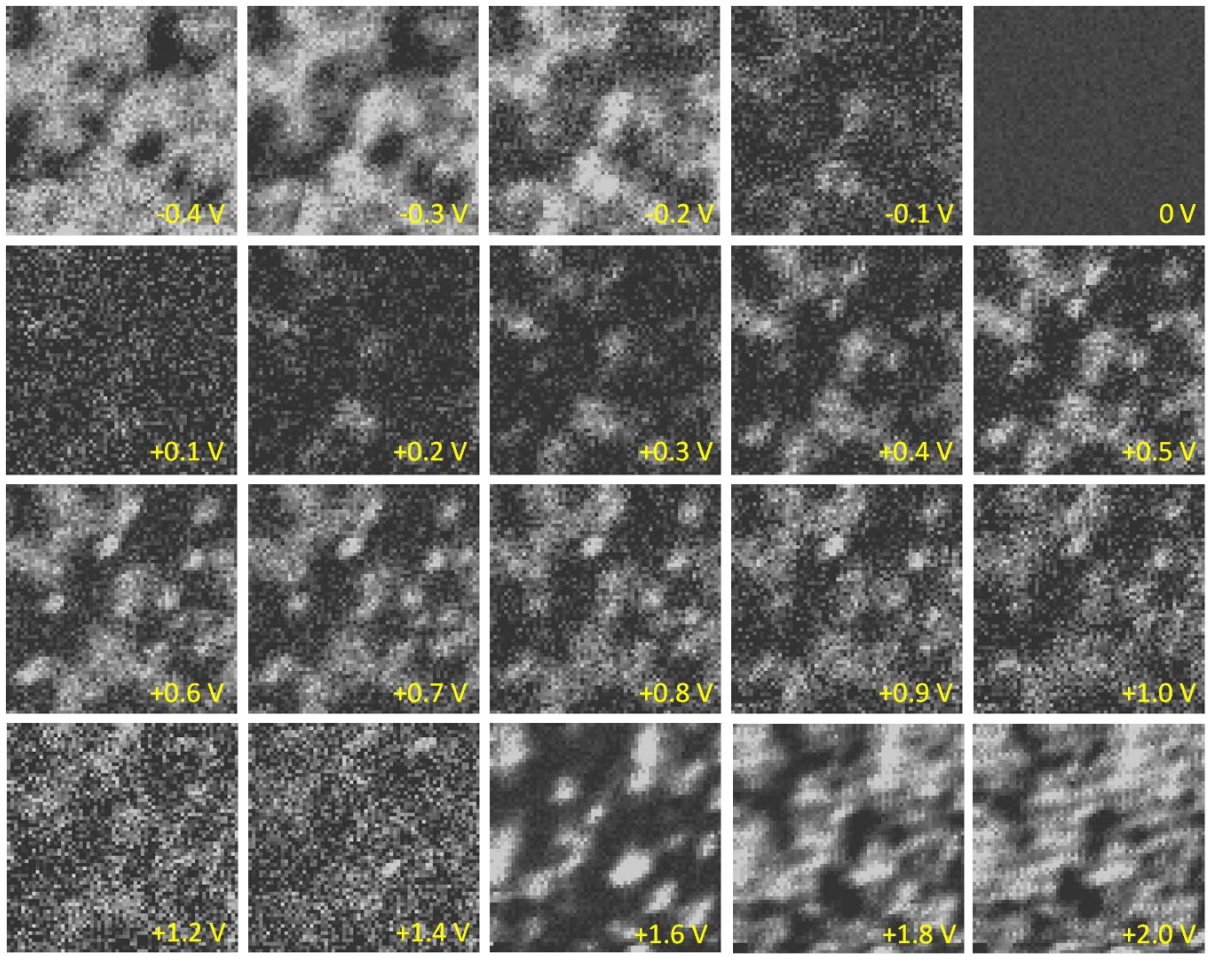

Fig. S2: Detailed *dI/dV* maps of the *p* = 0.03 sample measured in large energy window. The data are taken with tunneling junction set by *V* = -1.2 V and *I* = 4 pA. The *dI/dV* raw data is linearly scaled to brightness for each map.

# IV. Detailed DOS maps of the $p = 0.07$ sample

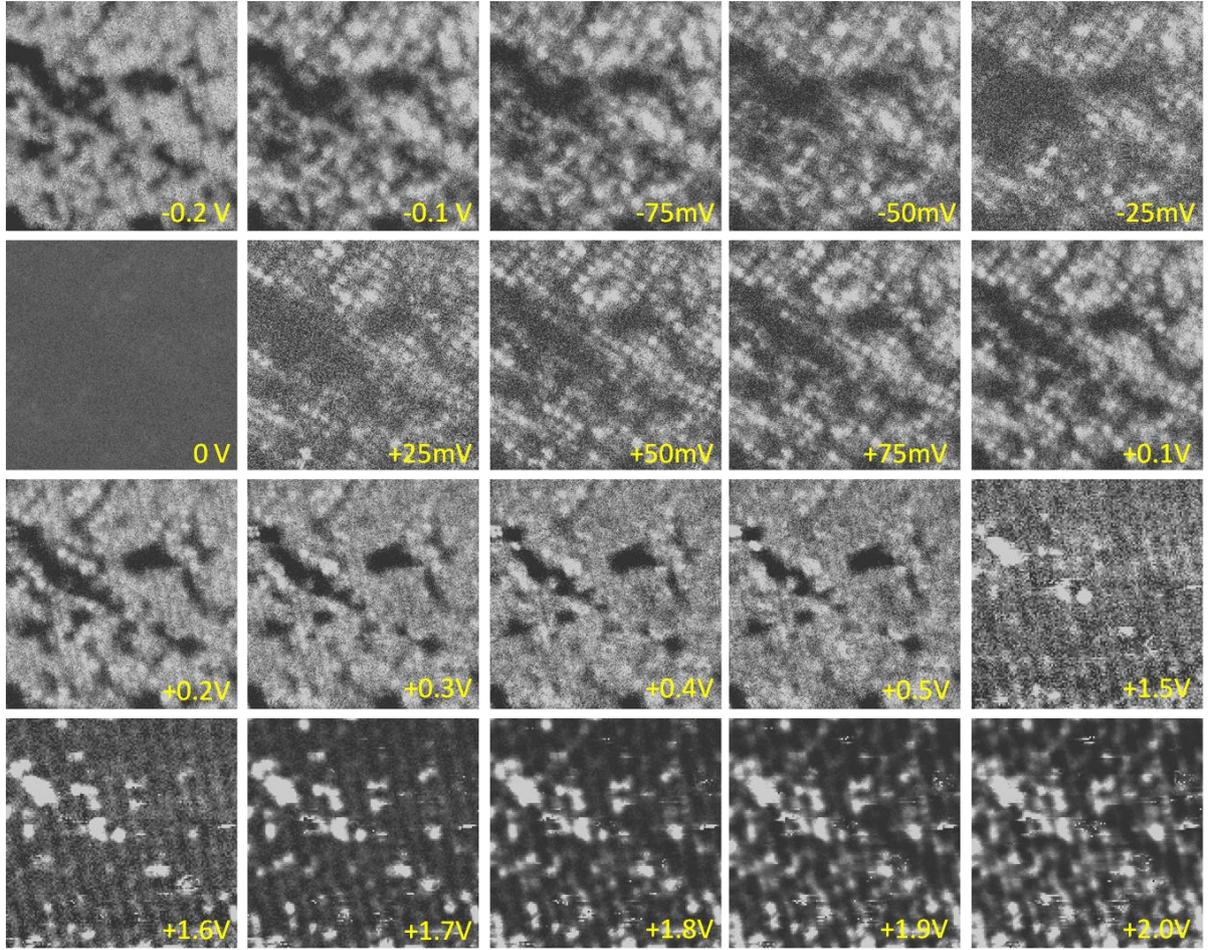

Fig. S3: Raw data of $dI/dV$ maps of the $p = 0.07$ sample measured over a large bias range. The maps for biases above +1.5 V are 128×128 pixel images taken with setup parameters $V = -1.2$ V and $I = 4$ pA. To have better energy and spatial resolutions at the low energy range, the maps for biases from -0.2 V to +0.5 V are 256×256 pixel images taken on the same area with setup parameters $V = -0.5$ V and $I = 4$ pA.

In Fig. 3 of the main text we display five representative DOS maps of the $p = 0.07$ samples with sample bias $V = -25$ mV, 0, +25 mV, +50 mV, and 1.7 V respectively. In Fig. S3 we show detailed raw data (in gray scale) of the DOS maps with more bias voltages taken on the same area of the $p = 0.07$ sample. The checkerboard pattern is most pronounced in the bias range from +25 mV to +100 mV, which is consistent with that observed on Bi-2212 at higher dopings (*4*). However, unlike the previous work, the checkerboard can also be observed on the negative bias side, such as − 25 mV and − 50 mV, although the features are weaker than that on the positive bias side. We note that superconducting La-Bi2201 with higher doping also show checkerboard pattern in both positive and negative bias sides (*5*). A closer look into their data also reads the checkerboard features are also clearer on the positive bias side. The strong resemblance of checkerboard features between insulating (here) and underdoped superconducting samples, including wavevector and energies dependence, underlies a similar origin of checkerboard at different doping regime in cuprates. The slight discrepancy between La-Bi2201 and Bi2212 in energies might be attributed to bilayer band splitting. Moreover, the positive correlation between the low bias maps (from -0.2 V to +0.5 V) and the anticorrelation with the high bias ones (above 1.5 V) are also present in this sample. Above +1.5 V, the structural supermodulation also starts to appear, just like in the $p = 0.03$ sample.

## V. Checkerboard pattern observed on another $p = 0.07$ sample

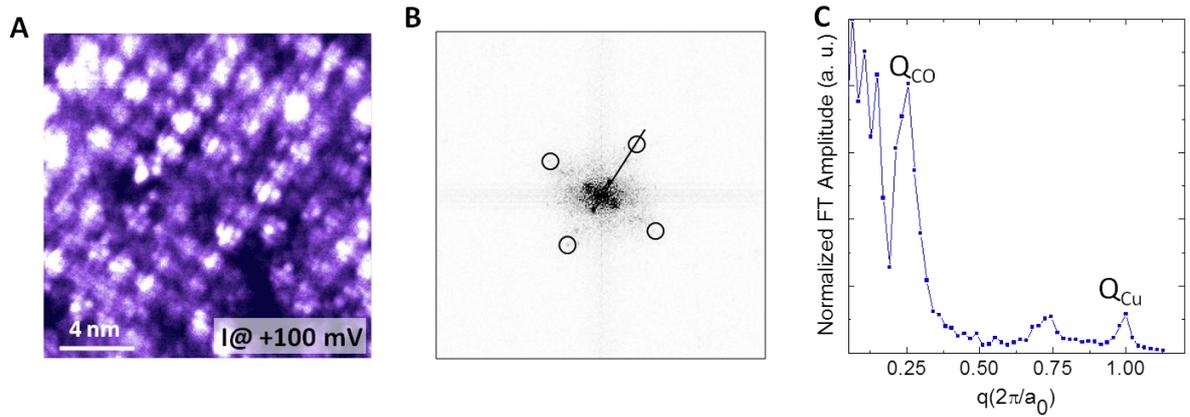

Fig. S4: (A) Current map taken at +100 mV on another piece $p = 0.07$ crystal from the same batch as that shown in the main text. The setup parameters are $V = -0.3$ V and $I = 4$ pA. (B) Fourier transform of the current map shown in (A), which shows the wavevectors for both the atomic lattice and the checkerboard. (C) The line-cut in the Fourier transform along the black line (Cu-O bond direction) shows a clear charge order peak at around 1/4 of the lattice wavevector, indicating a real-space checkerboard periodicity of $4a_0$.

To check if the checkerboard pattern observed on the $p = 0.07$ sample is reproducible, we have performed the same experiment on another piece of La-Bi2201 crystal from the same batch. To exclude the possibility of artifacts due to the STM tip or setup parameters, we have retreated and recalibrated the tip *in situ*. The setup parameters for the *dI/dV* map are also changed to $V = -0.3$ V and $I = 4$ pA. Shown in Fig. S4A is the current map taken at bias voltage +100 mV, which clearly reveals the checkerboard pattern that looks the same as that shown in Fig. 4A in the main text. Fig. S4B and S4C are the FFT image of the current map and the linecut along the Cu-O bond direction. The wavevector of the checkerboard is also at 1/4 of the lattice wavevector $2\pi/a_0$, indicating a checkerboard along the Cu-O bond direction with periodicity $4a_0$. All the checkerboard features shown in the main text are thus confirmed.